\documentclass{edbk} 

\usepackage{epsfig}
\normallatexbib
\begin{document}
\articletitle[Two-proton correlations at AGS]      
{Beam energy dependence of two-proton correlations at the AGS}  
\author{Sergei Y. Panitkin,$^7$ for the E895 Collaboration}
\email{panitkin@sseos.lbl.gov}
\author{N.N.~Ajitanand,$^{12}$ J.~Alexander,$^{12}$ M.~Anderson,$^5$ 
D.~Best,$^1$ F.P.~Brady,$^5$ T.~Case,$^1$ W.~Caskey,$^5$ 
D.~Cebra,$^5$ J.~Chance,$^5$ P.~Chung,$^{12}$
B.~Cole,$^4$ K.~Crowe,$^1$ A.~Das,$^{10}$
J.~Draper,$^5$ M.~Gilkes,$^{11}$ S.~Gushue,$^2$
M.~Heffner,$^5$ A.~Hirsch,$^{11}$
E.~Hjort,$^{11}$ L.~Huo,$^6$ M.~Justice,$^7$
M.~Kaplan,$^3$ D.~Keane,$^7$ J. Kintner,$^8$
J.~Klay,$^5$ D.~Krofcheck,$^9$ R.~Lacey,$^{12}$ M.~Lisa,$^{10}$
H.~Liu,$^7$ Y.~Liu,$^6$ R.~McGrath,$^{12}$ Z.~Milosevich,$^3$, G.~Odyniec,$^1$
D.~Olson,$^1$  C.~Pinkenburg,$^{12}$
N.~Porile,$^{11}$ G.~Rai,$^1$ H.-G.~Ritter,$^1$
J.~Romero,$^5$ R.~Scharenberg,$^{11}$ L.~Schroeder,$^1$
B.~Srivastava,$^{11}$ N.~Stone,$^2$ T.J.M.~Symons,$^1$ 
S.~Wang,$^7$ J.~Whitfield,$^3$ T.~Wienold,$^1$ R. Witt,$^7$ 
L.~Wood,$^5$ X.~Yang,$^4$ W.~Zhang,$^6$ Y.~Zhang$^4$\\}
\affil{
$^1$Lawrence Berkeley National Laboratory, Berkeley, California 94720\\
$^2$Brookhaven National Laboratory, Upton, New York 11973\\
$^3$Carnegie Mellon University, Pittsburgh, Pennsylvania 15213\\
$^4$Columbia University, New York, New York 10027\\
$^5$University of California, Davis, California 95616\\
$^6$Harbin Institute of Technology, Harbin 150001, P.~R.~China \\
$^7$Kent State University, Kent, Ohio 44242\\
$^8$St.~Mary's College of California, Moraga, California 94575\\
$^9$University of Auckland, Auckland, New Zealand \\
$^{10}$The Ohio State University, Columbus, Ohio 43210\\
$^{11}$Purdue University, West Lafayette, Indiana 47907\\
$^{12}$State University of New York, Stony Brook, New York 11794\\
}


\begin{abstract}
First measurements of the beam energy dependence of the two proton
correlation function in central Au+Au collisions are performed by
the E895 Collaboration at the BNL AGS. No significant changes with
beam energy were observed. The imaging technique of
Brown-Danielewicz is used in order to extract information about
the space-time content of the proton source at freeze-out. Extracted
source functions show peculiar enhancement at low relative separation.
\end{abstract}


\section{Introduction}
Two-particle correlations are widely considered to be a valuable tool
in extracting information about the space-time extent of the system
created in the collisions of heavy
ions~\cite{koonin_77,lednicky_82,gelbke_90,pratt_90}. The complex
nature of the  heavy-ion reaction requires utilization of different
particle species in order to obtain reliable and complete picture of
the system created in the collision. 
The majority of the existing experimental two-particle correlation data in
ultra-relativistic heavy-ion collisions was obtained using mesons
as a probe.
The data on baryon correlations is sparse at best. Since the
physics of heavy-ion collisions in the beam energy range between 1 and
11 AGeV is dominated by baryons and baryon resonances, the information related
to the space-time extent of the baryon source, obtained via two-proton
correlations is clearly very interesting.  In this paper we present 
preliminary results of the first measurement of the beam energy
dependence of the two-proton 
correlation function in the central Au+Au collisions at 2,4,6 and 8
AGeV performed by the E895 Collaboration at the Brookhaven National Lab
(BNL) Alternating Gradient Synchrotron (AGS). Preliminary results of
the pion correlation analysis were published elsewhere~\cite{lisa_1}.  
\section[E895]
{Experimental Details}
E895 is a fixed target experiment at the BNL AGS. 
The goal of E895 is to study multiparticle correlations and particle 
production with Au beams incident on a variety of targets, over a
range of AGS energies. 
More information about the E895 experimental setup can be found
elsewhere~\cite{rai_90, bauer_97}. We will describe in the
following only detailes relevant to the presented analysis.\\
\indent Beams  of gold ions 
($^{197}$Au) were available at different energies - 2,4,6 and 8
AGeV. They were used to bombard targets of different materials- Be,
Cu, Ag and Au. Charged particles produced in the collision were detected
with time projection chamber (TPC) ~\cite{rai_90}, positioned in side
the MPS magnet, and multi-sampling ionization chamber
(MUSIC)~\cite{bauer_97} located downstream from the magnet. For the  
presented results only information from the TPC was
used. The time projection chamber is filled with P10 gas and has
rectangular fiducial volume which is about 150 $cm$ long, 75 $cm$ high,
and 100 $cm$ wide. The ionization produced by charged particles in the
chamber is detected by 
a segmented cathode plane at the bottom of the TPC. The cathod plane
has 15360(120 by 128) pads. The dimension of the pads are 0.8 $cm$ by
1.2 $cm$. The signal from each pad is sampled 140 times at 10 MHz by a
12 bit flash ADC yielding more than 2 millions pixels per event used
for track reconstruction. The TPC was capable of detecting and
tracking software of reconstructing up to several hundreds tracks per 
event. The magnetic field of the MPS magnet was typically 0.75  
or 1 Tesla. Particle identification was performed via simultaneous
measurement of particle momentum and specific ionization in the TPC
gas. It was possible to resolve positively charged particles up to
charge 6 and obtain reliable identification of protons up to 0.9 GeV/c
in momentum. 
\section{Data analysis}
Good momentum resolution and good particle identification 
capabilities together with high reconstructed charged particle
multiplicity allowed the performance of two-particle correlation
studies.
In order to obtain the two-proton correlation function $C_{2}$
experimentally, the mixed event technique was used. We employ the
following definition of the correlation function
\begin{eqnarray} C_2(q_{inv}) =
\frac{N_{tr}(q_{inv})}{N_{bk}(q_{inv})} \hspace{0.3cm}, \end{eqnarray}
\noindent{where}                                            
\begin{eqnarray}                                                     
q_{inv} = q = \frac{1}{2} \sqrt{-(p_1^{\mu} - p_2^{\mu})^2}              
\end{eqnarray}                                                      
\noindent{is the half relative invariant momentum between the two
identical    
particles with four-momenta $p_1^{\mu}$ and $p_2^{\mu}$. The
quantities  N$_{tr}$ and N$_{bk}$
are the ``true'' and ``background'' two-particle distributions
obtained by selecting particles from the same and different events,     
respectively. Before calculating the correlation function, several
cuts are applied. In order to insure a reliable particle
identification and high purity of the proton sample, a cut on proton
longitudinal momentum $P_z<800$ MeV/c is applied.
Contamination of the identified proton sample by other particles, in
this momentum interval, was estimated to be less than $2\%$.  Event
centrality selection is based on a reconstructed 
charged particles multiplicity. For the present analysis events are
selected with a multiplicity cut corresponding to the upper $5\%$ of
the inelastic cross section for the Au+Au collisions. Single proton
tracks are required to satisfy certain quality cuts. Number of hits
belonging to the track should be greater than 20, thus suppressing
short tracks from delta electrons and remnants of the split
tracks. Track should point into vicinity of the event vertex with
distance of closest approach (DCA) less than 2.5 cm. Tracks should be
properly reconstructed by the tracking code with the corresponding
$\chi^2$ per degree of freedom less than 1.5. Tracks are required to
be reconstructed from fairly continuous sequence of hits with no
significant hit losses, the fraction of hits assigned to the track
should exceed 50$\%$ of the theoretically available number for the
corresponding trajectory inside the fiducial volume of the TPC. This
cut has been shown to be effective in suppression of the track
splitting effects in the correlation analysis~\cite{lisa_1}.
In order to suppress effects of track merging, a cut on angular
separation of two tracks was imposed. For pairs from 
both ``true'' and ``background'' distributions the angle between
two tracks was required to be greater than 3 degrees.
Figure~\ref{panitkin_1} shows measured two-proton correlation
functions for Au+Au central collisions at 2,4,6 and 8 AGeV. Within
currently available statistical accuracy no significant changes of the
measured correlation functions with beam energy were observed. 
\begin{figure}[ht]
\vskip40pt
\epsfig{file=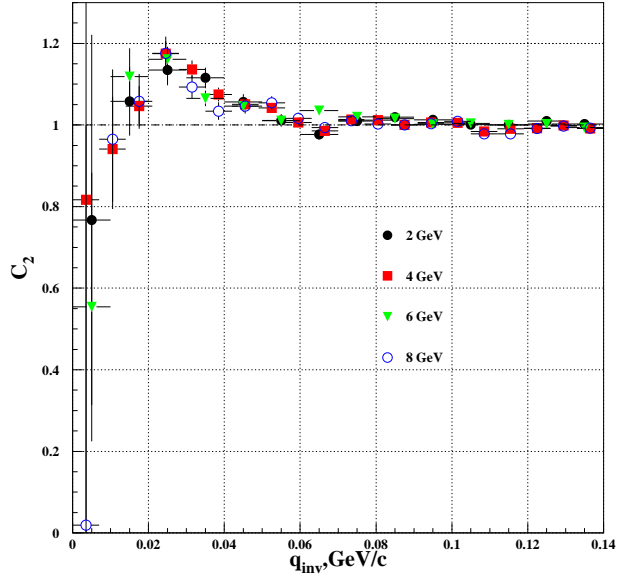,height=9cm,width=9cm}
\caption{Measured two-proton correlation functions for Au+Au central
collisions at different beam energies. } 
\label{panitkin_1}
\end{figure}
\section{Source Imaging}\inxx{Source Imaging}
The so called source imaging technique of Brown-Danielewicz was used
to extract information about the space-time extent of the proton
source. Here we will give just a brief
sketch of the method, see Refs~\cite{dbrown_1,
dbrown_2}, for a more detailed description.
The two-particle correlation function may be expressed in the
following way:
\begin{equation}
C_{\bf P}({\bf q}) = {d N_2 / d {\bf p}_1 \, d {\bf p}_2
\over \left( d N_1 / d {\bf p}_1 \right)
\left( d N_1 / d {\bf p}_2 \right) }
 \simeq
\int d{\bf
r} \, |\Phi_{\bf q}^{(-)}({\bf r})|^2 \, S_{\bf P} ({\bf r}).
\label{panitkin_CPq}
\end{equation}
$S_{\bf P} ({\bf r})$ is the distribution of relative separation 
of emission points for the two particles, in their center of mass and
$\Phi_{{\bf q}}^{(-)}({\bf r})$ is a relative wave function. 
Using single-particle sources,
\begin{equation}
S_{\bf P} ({\bf r}) = \int d{\bf R} \, dt_1 \, dt_2 \,
\overline{D}(0,{\bf R} + {\bf r}/2, t_1) \,
\overline{D}(0,{\bf R} - {\bf r}/2, t_2) \, .
\label{panitkin_Spr}
\end{equation}
where $\overline{D}$ is an averaged distribution of freeze-out points
of the particles. 
In~the proton-proton case, the~angle and spin averaged relative
wave function can be expressed as
\begin{equation}
|\Phi_{{\bf q}}^{(-)}({\bf r})|^2 = {1 \over 2} \sum_{j s \ell \ell'} (2 j +1) \
\left( g_{js}^{\ell \ell'} (r) \right)^2  \, ,
\end{equation}
where $g_{js}^{\ell \ell'}$ is the radial wave function
with outgoing asymptotic
angular momentum~$\ell$, which can be calculated numerically given a
particular description of the final state interaction. In the present
analysis, the proton relative wave functions were calculated by solving
the  Schr\"odinger equation with the  REID93\cite{sto94} and Coulomb 
potentials.
The imaging method is concerned with the determination of the relative
source function ($S_{\bf P} ({\bf r})$ in Eq.~\ref{panitkin_CPq}) knowing $C_{\bf
P}({\bf q})$. Taking into account that the nontrivial part of the
correlation function is deviation from unity, one may rewrite
Eq.~\ref{panitkin_CPq} in the following way
\begin{equation}
C_{\bf P}({\bf q}) -1 =
\int d{\bf
r} \,
\left(|\Phi_{\bf q}^{(-)}({\bf r})|^2 -1 \right) \,
S_{\bf P} ({\bf r}) =
\int d{\bf
r} \,
K({\bf q}, {\bf r}) \, S_{\bf P} ({\bf r}) ,
\label{panitkin_K}
\end{equation}
where $K = |\Phi_{\bf q}^{(-)}|^2-1$.
The~problem of imaging then reduces to the
more general problem of inversion~\cite{tarantola} of~$K$ in
Eq.~\ref{panitkin_K}.\\
\indent Figure~\ref{panitkin_2} shows the relative distribution of
emission points of protons for central Au+Au collisions at 2,4,6 and 8
AGeV obtained as a result of the application of the imaging technique
described above. In order to check the quality of the imaging and
numerical stability of the inversion procedure the 
two-proton correlation functions are calculated using the 
relative source functions shown on Figure~\ref{panitkin_2} as an input
in Equation~\ref{panitkin_CPq}. The result of such ``double inversion''
procedure is shown on Figure~\ref{panitkin_3} for the beam energy 4
AGeV. The agreement between the measured and reconstructed correlation
function is quite good.\\ 
\indent It can be seen from Figure~\ref{panitkin_2} that the relative
proton source functions have similar shapes at all measured
energies. Extracted source functions 
show enhancement at low relative separation which may be induced
by momentum position correlations in the source, possibly due to
collective flow. Further investigation and understanding of the origin
of this enhancement is clearly needed. Even though the source
functions have a  non-trivial overall shape, the tail of the relative
source function may be fit by gaussian(\ref{panitkin_gauss}) or
exponential (\ref{panitkin_expo}) forms:   
\begin{equation}
S(r)=\frac{\lambda}{(2{\pi}R^2)^{3/2}}exp(-r^2/2R^2)
\label{panitkin_gauss}
\end{equation}
\begin{equation}
 S(r)=\frac{\lambda}{2R^3}exp(-r/R)
\label{panitkin_expo}
\end{equation}
Parameter $\lambda$ is a so called generalized chaoticity
parameter~\cite{dbrown_1} defined as: 
 \begin{equation}
\lambda (r_N) = \int_{r < r_N} d{\bf r} \, S({\bf r})
\end{equation}
and has the meaning of an integral of the source over a region ($r <
r_N$) where the distribution is significant. 
Results of the fit of the tail of the correlation function using these
two parameterizations are shown at Figure~\ref{panitkin_4}. Fit
parameters are presented in Table~\ref{panitkin_t_1}. At the current
level of precision of the data both parameterizations provide an
adequate description of the relative source function, except for the
separations smaller than 2 fm. 
\begin{figure}[ht]
\vskip4pt
\epsfig{file=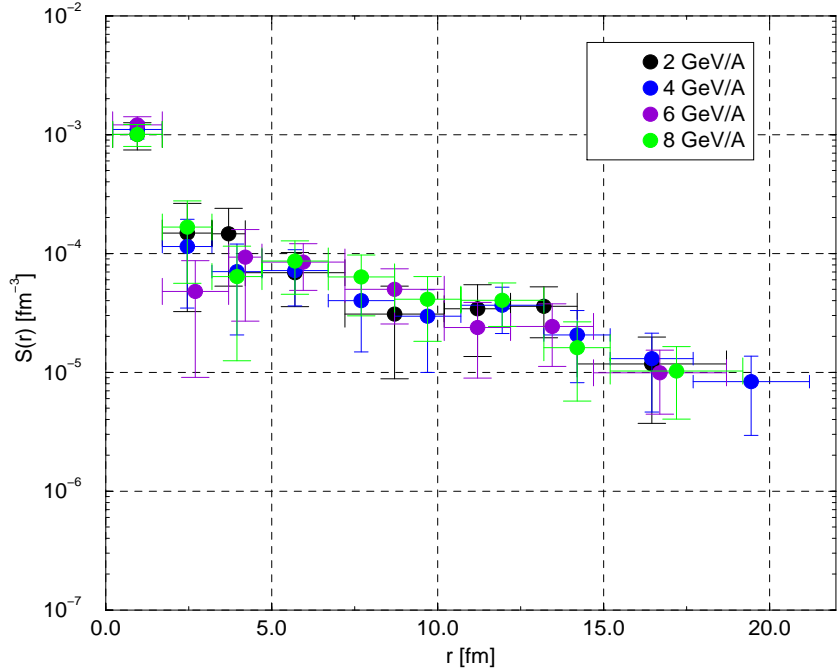,height=11cm,width=9cm, angle=270}
\caption{Relative source functions extracted from the proton
correlation data at 2,4,6 and 8 AGeV. } 
\label{panitkin_2}
\end{figure}
\begin{figure}[ht]
\vskip4pt
\epsfig{file=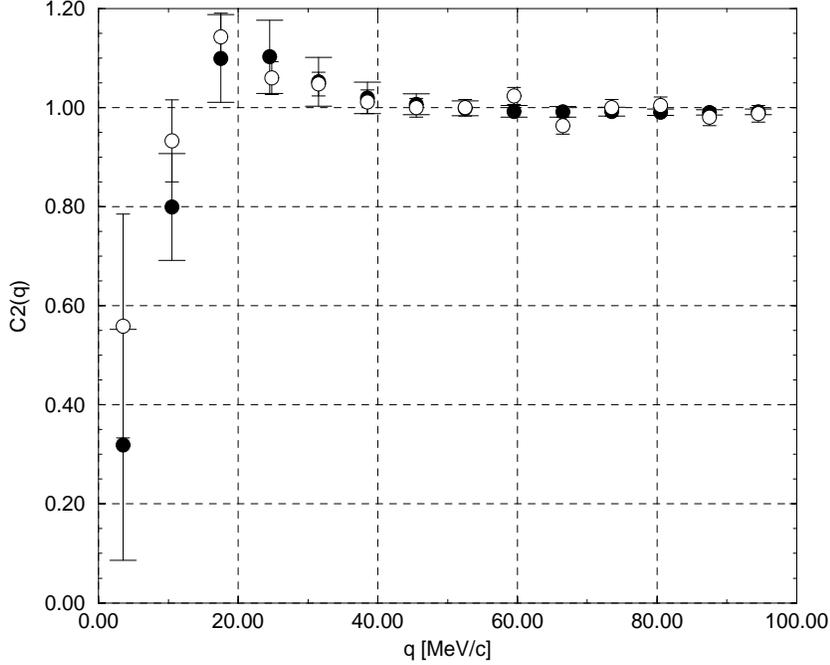,height=11cm,width=9cm, angle=270}
\caption{Experimentally measured two-proton correlation function (open
circles) and correlation function restored from the relative source
(filled circles) for beam energy 4 AGeV. See description in the text.}  
\label{panitkin_3}
\end{figure}
\begin{figure}[ht]
\vskip4pt
\epsfig{file=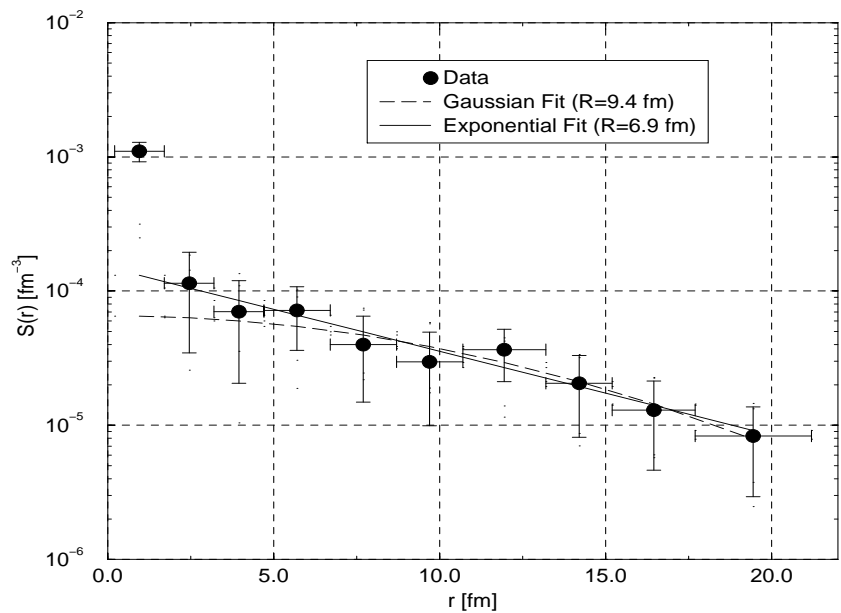,height=11cm,width=9cm, angle=270}
\caption{Gaussian and exponential fits to the relative source
reconstructed for the beam energy 4 AGeV.}  
\label{panitkin_4}
\end{figure}
\begin{table}
\caption{Fit parameters for gaussian and exponential
parameterizations of the relative source functions for different beam 
energies. See description in the text.} 
\begin{tabular*}{\textwidth}{@{\extracolsep{\fill}}cccccc}
 \hline
 $E_{b}$ (AGeV)&$\lambda$ Exp.& R Exp.(fm)&$\lambda$ Gauss.& R
 Gauss.(fm)&$R_{N}$ (fm)\\ \hline  
  {{2.0}}&{{1.37}}&{{6.85}}&{{0.98}}&{{9.00}}&{{18.7}}\\ 
  {{4.0}}&{{1.27}}&{{6.95}}&{{0.87}}&{{9.45}}&{{21.2}}\\
  {{6.0}}&{{0.97}}&{{5.61}}&{{0.69}}&{{7.88}}&{{18.7}}\\
  {{8.0}}&{{1.10}}&{{5.66}}&{{0.79}}&{{7.89}}&{{19.2}}\\
\hline 
\end{tabular*}
\label{panitkin_t_1}
\end{table}
\section{Summary}
We reported preliminary results of the analysis of the beam energy
dependence of the two-proton correlation function in the target
fragmentation region ($P<800$ MeV/c). The correlation
functions were measured for the first time for protons in central
Au+Au collisions at beam energies 2,4,6 and 8 AGeV. Within currently
available statistical accuracy no significant changes with beam energy
were observed. The source imaging technique of Brown-Danielewicz was used
to extract information about the space-time extent of the proton
source. It was found that the relative proton source functions have
similar shapes at all measured energies. Extracted source functions
show enhancement at low relative separation which may be induced
by momentum-position correlations in the source, possibly due to
collective flow. Further investigation of the origin of this
enhancement is clearly needed.
\begin{acknowledgments}
Stimulating discussions with Drs.~N.Xu and S.~Voloshin are gratefully acknowledged.
The author wish to thank P.~Danielewicz and D.~Brown for performing
source imaging calculations. This research is supported by the
U.S. Department of Energy , the U.S. National Science Foundation and
by University of Auckland, New Zealand, Research Committee.
\end{acknowledgments}

\end{document}